\documentclass[twocolumn,groupedaddress,
superscriptaddress,
showpacs,preprintnumbers,
amsmath,amssymb,
aps,
]{revtex4}

\usepackage{graphicx}
\usepackage{dcolumn}
\usepackage{bm}%
\usepackage[colorlinks=true,citecolor=blue]{hyperref}
\hypersetup{colorlinks=true,citecolor=blue,linkcolor=red,urlcolor=blue}

\usepackage{float}

\usepackage{color}
\definecolor{Green}{RGB}{0,204,102}
\definecolor{Purple}{RGB}{102,0,255}
\definecolor{Blue}{RGB}{51,153,255}
\definecolor{Red}{RGB}{151,010,010}

\begin{document}

\sloppy

\title{Low-loss two-dimensional plasmon modes in antimonene}

\author{Zahra Torbatian}
\affiliation{School of Nano Science, Institute for Research in Fundamental Sciences (IPM), Tehran 19395-5531, Iran}
\author{Mohammad Alidoosti}
\affiliation{School of Nano Science, Institute for Research in Fundamental Sciences (IPM), Tehran 19395-5531, Iran}
\author{Dino Novko}
\affiliation{Institute of Physics, Bijeni\v{c}ka 46, 10000 Zagreb, Croatia}
\affiliation{Donostia International Physics Center (DIPC),
Paseo Manuel de Lardizabal 4, 20018 Donostia-San Sebasti\'an, Spain}
\author{Reza Asgari}
\email{asgari@ipm.ir}
\affiliation{School of Nano Science, Institute for Research in Fundamental Sciences (IPM), Tehran 19395-5531, Iran}
\affiliation{School of Physics, Institute for Research in Fundamental Sciences (IPM), Tehran 19395-5531, Iran}
\affiliation{ARC Centre of Excellence in Future Low-Energy Electronics Technologies, UNSW Node, Sydney 2052, Australia}

\begin{abstract}
The effects of spin-orbit (SOC) and electron-phonon coupling on the collective excitation of doped monolayer Sb$_2$ are investigated using density functional and many-body perturbation theories. The spin-orbit coupling is exclusively important for the monolayer Sb$_2$ and it leads to the reconstruction of the electronic band structure. In particular, plasmon modes of monolayer Sb$_2$ are quite sensitive to the SOC and are characterized by very low damping rates owing to small electron-phonon scatterings. Our results show plasmons in antimonene are significantly less damped compared to monolayer graphene when plasmon energies are $\hbar \omega> 0.2$\,eV due to smaller plasmon-phonon coupling in the former material.  
\end{abstract}

\pacs{73.20.Mf, 71.10.Ca, 71.15.-m, 78.67.Wj}
\maketitle

\section{Introduction}\label{sec:intro}
There has been a large amount of experimental and theoretical activities in recent years to explore the optical and transport properties of two-dimensional (2D) crystalline materials.
Advances in fabrication techniques have made it possible to probe various quantities of interest in high quality and various charge density samples. Graphene was the first 2D crystalline material to be isolated
in 2004~\cite{novoselov2004electric} and there have been literally hundreds of other materials, with a vast range of properties~\cite{geim2013van, mounet2018two}.
Antimonene, a single layer antimony and another member of the nitrogen group, is a recently discovered 2D semiconductor with exceptional environmental stability\,\cite{stability, ares2018recent, ji2016two,wu2017epitaxial, shi2019van}, potentially possessing high carrier mobility\,\cite{high_mobility} and strong spin-orbit coupling (SOC)\,\cite{kurpas2019spin, rudenko2017}. Moreover, it has been isolated both by mechanical\,\cite{Ares2016,lloret2019few} and liquid-phase exfoliations\,\cite{Gibaja2016}.
Antimonene has several crystal phases,  such as $\alpha$, $\beta, \gamma$ and $\delta$.  Among them, the $\beta$ phase shows the lowest Free-energy state~\cite{zhang2016semiconducting}.
In contrast to puckered phosphorene, the $\beta$-Sb
holds buckled honeycomb structure with much stronger spin-orbit coupling (SOC), which brings exotic fundamental properties for photonics and
spintronics\,\cite{zhang2017}. A less stable structure is a puckered $\alpha$-Sb with two atomic sub-layers. Monolayer $\beta$-Sb is a semiconductor with an indirect band gap and it is thus suitable for application in optoelectronics, where thickness- and strain-tunable band gap could provide additional control over the materials properties\,\cite{Lugovskoi2019}.
Furthermore, antimonene was shown to display remarkable optical and electronic properties that could be managed and tuned by applied strain and external electric fields\,\cite{shu2018}.

Owing to these special electronic structure properties, antimonene could support exceptional collective charge excitations, i.e., plasmons,~\cite{torbatian2018plasmonic} suitable for application in plasmonics and nanophotonics\,\cite{Prishchenko2018,katsnelson2018}. Namely, by using the tight-binding model it was recently shown that under application of the gate voltage electron-doped antimonene demonstrates unusual low-loss plasmonic excitations in the mid-infrared region\,\cite{Prishchenko2018}. However, the phonon-assisted electron processes that are instrumental for determining plasmon decay rates below Landau (interband) damping region have not been considered thus far. For instance, these processes (i.e., couplings with intrinsic acoustic and optical phonons) were shown to dictate plasmon lifetimes in high-mobility graphene\,\cite{principi14,novko17,ni18}. Therefore, in order to fully characterize plasmonic excitations in antimonene, it is important to consider these electron-phonon coupling (EPC) effects.

In this work, we investigate the electronic and optical properties of electron- and hole-doped antimonene in its most stable $\beta$-phase using first-principles calculations. We consider the role of the SOC in the electronic structure and plasmon dispersions. A wide optical absorption in the energy range of $2-4$ eV is obtained in the system. Having noted, the SOC causes a reconstruction of the band structure and induces a splitting of the valence band. As a consequence, the plasmon spectra of the doped antimonene is highly affected by the SOC-induced band structure modifications. Moreover, a particular attention might be devoted to electron-phonon interactions, since phonons play a dominant role in suppressing the intrinsic mobility of a material. In order to have a realistic electronic structure, we consider the effect of charge carrier doping effects by making use of a jellium model. We show that the impact of the EPC on plasmon decay rates dispersion is favourably small in a doped antimonene, which makes this novel 2D material promising and appealing for applications in plasmonics.  

This paper is organized as follows. We begin with a description of our theoretical formalism in Sec. II, followed by the details of the DFT simulations, optical formalism in the presence of the EPC. Numerical results of the absorption, rigid band approximation, charge doping in the jellium model, charge plasmon modes and lifetime of the phonon scattering process together with the Kohn anomaly descriptions are presented in Sec. III. We summarize our main findings in Sec. IV.

\section{THEORY AND COMPUTATIONAL METHODS}\label{sec:theo}

We investigate the electronic and optical properties of antimonene with the presence of the SOC and EPC interactions. To this aim, we use a combination of density functional theory (DFT) and density functional perturbation theory (DFPT)~\cite{baroni2001phonons} in conjunction with the formalism of maximally localized Wannier functions~\cite{mostofi2008wannier90} as implemented in EPW code~\cite{ponce2016epw}.
The {\it ab-initio} calculations are carried out in the framework of the local density approximation (LDA) of the density functional theory and norm-conserving pseudopotential within the QUANTUM ESPRESSO package\cite{0953-8984-21-39-395502}, using Perdew-Zunger LDA exchange-correlation.

\begin{figure*}[!t]
\begin{center}
\includegraphics*[width=16.4cm]{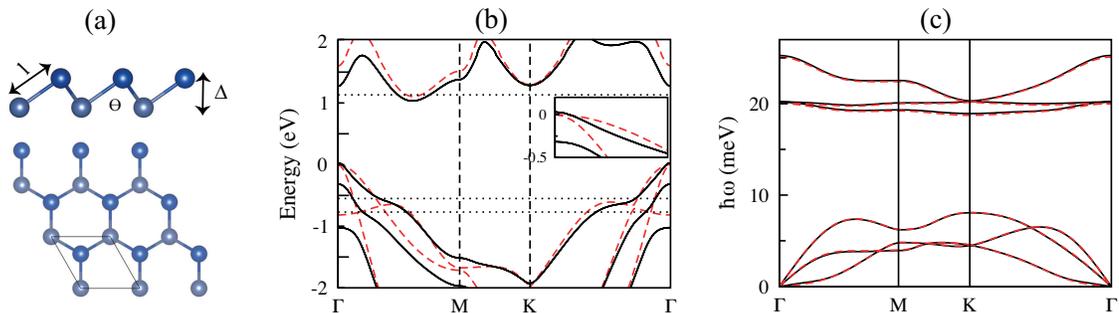}
\caption{(Color online) (a) Side and top view of monolayer $\beta$-Sb. $l$, $\Delta$ and $\theta$ are bond length, buckling height and bond angle, respectively. (b) Electronic band structures and (c) phonon spectrum of monolayer $\beta$-Sb along the high symmetry $\Gamma$-M-K-$\Gamma$ points with (black line) and without (red dashed-line) spin-orbit coupling. The values of the Fermi energy for different values of doping are shown by black dotted lines and those are 0.2 and 0.1 holes/u.c. and 0.1 electrons/u.c. from bottom to top, respectively.}
\label{fig1}
\end{center}
\end{figure*}

We use kinetic cutoff energy of $50$\, Ry and a vacuum spacing of about $25\,\textrm{\AA}$ to avoid the effects of the interaction between periodic images in the $z$ direction. The convergence criterion for energy is set to $10^{-8}$ eV and the atomic positions are relaxed until the Hellmann-Feynman forces are less than $10^{-4}$ eV/\AA. A set of $30 \times 30 \times 1$ ~$\Gamma$-centered $k$-point sampling is used for the ground-state calculations.

\subsection{Optical absorption spectra and phonon-induced damping}\label{method}
We would like to explore the optical absorption of the system by making use of the current-current response tensor calculated within DFT where the electromagnetic interaction is mediated by the free-photon propagator. To this aim, we pursue the same procedure given in Refs.\,\cite{Novko2016,torbatian_PRB,rukelj2016optical,despoja2009propagators} to obtain the optical properties of antimonene.
First, we consider independent electrons which live in a local crystal potential obtained by DFT and interact with the electromagnetic field described by the vector potential. Then, we solve the Dyson equation for the screened current-current response tensor in the quasi-2D crystal of one or few layers
$\Pi=\Pi^0 + \Pi^0 \otimes D^0 \otimes \Pi$,
where $\Pi^0$ and $D^0$ are the non-interacting current-current response tensor and free-photon propagator, respectively.

The non-interacting current-current response tensor can be written as

\begin{eqnarray}
\Pi^0_{\mu\nu, G_z, G'_z}(\mathbf q, \omega)= \frac{1}{V} \sum_{\mathbf k, n, m} \frac{\hbar \omega}{E_n(\mathbf k)-E_m(\mathbf k+\mathbf q)}\nonumber\\
\times [J^\nu_{\mathbf k n , \mathbf k+\mathbf q m} (G'_z)]^* J^\mu_{\mathbf k n , \mathbf k+\mathbf q m} (G_z)\nonumber\\
\times\frac{f_n(\mathbf k)-f_m(\mathbf k+\mathbf q)}{\hbar\omega + i\eta+ E_n(\mathbf k) - E_m(\mathbf k+\mathbf q)}\nonumber\\
\label{current}
\end{eqnarray}
where $J^\mu_{\mathbf k n , \mathbf k+\mathbf q m} (G_z)$ are the current vertices (see Refs.\,\cite{Novko2016, torbatian_PRB} for more details) and $E_n(\mathbf k)$ are the Kohn-Sham energies.
Here $f_n(\mathbf k) $ is the Fermi-Dirac distribution at temperature $T$, $G_z$ is reciprocal lattice vector along the perpendicular of the system and $V$ is the normalized volume. Notice that we define three-dimensional vector ${\bf r}=({\bf \rho}, z)$ and the Kohn-Sham wave functions are expanded over the plane waves with coefficients that are obtained by solving the Kohn-Sham equations self-consistently. Further, the summation over $\mathbf k$ wavevectors is carried on a 101$\times$101$\times$1 grid, $n$ index sums over $20$ electronic bands, and  polarization direction is $\mu, \nu=x, y, z$.
Finally, the optical absorption is given by $A(\mathbf q,\omega)=-4\hbar  {\rm Im}\,\Pi_{\mu\mu}(\mathbf{q},\omega)/\omega$\,\cite{Novko2016,novko17,torbatian_PRB}.

In order to investigate the effects of phonon on the plasmon dispersion, we use the formalism that was presented in Ref.\,\cite{novko17,Caruso2018}. Optical excitations are first convenient to decompose into the intraband ($n=m$) and interband ($n\ne m$) contributions. The electron-phonon scattering mechanism is then considered in the intraband channel.

For ${\mathbf q}\approx 0$, the intraband contribution of current-current response tensor can be written as the following~\cite{PhysRevB.3.305}:

\begin{eqnarray}
\Pi^0_{\mu\mu}= \frac{2}{V} \frac{\omega}{\omega[1+\lambda_{\rm ph}(\omega)]+i/\tau_{\rm ph}(\omega)}\sum_{\mathbf k, n} \frac{\partial f_{nk}}{\partial E_{n\mathbf k}}  |J^\mu_{nn\mathbf k}|^2
\label{intra_cu}
\end{eqnarray}
Here the effects of the EPC are contained in the scattering time and the dynamical renormalization parameters, i.e., $\tau_{\rm ph}(\omega)$ and $\lambda_{\rm ph}(\omega)$, respectively, which are defined as

\begin{eqnarray}
\tau_{\rm ph}^{-1}(\omega)=\frac{2\pi\hbar}{\omega} \int_{0}^{\omega} d\Omega (\omega-\Omega)\alpha^2F(\Omega),
\label{tau}
\end{eqnarray}
and
\begin{eqnarray}
\lambda_{\rm ph}(\omega)&=&-\frac{2}{\omega} \int_{0}^{\infty} d\Omega \alpha^2F(\Omega)\nonumber\\
 &&\times\Big[\ln\Big|\frac{\omega-\Omega}{\omega+\Omega}\Big|-\frac{\Omega}{\omega} \ln\Big|\frac{\omega^2-{\Omega}^2}{{\Omega}^2}\Big|\Big],
\label{landa}
\end{eqnarray}
where $\alpha^2F(\Omega)$ is the Eliashberg spectral function\,\cite{novko17,Caruso2018}.
It has been shown that this method is an extension of the Drude model to account the electron-phonon interaction with the additional of a frequency dependence for the scattering rate~\cite{PhysRevB.3.305}. In other words, in order to fulfill the causality, the effective mass renormalization $1+\lambda_{\rm ph}(\omega)$ also becomes frequency dependent~\cite{Puchkov_1996}.

\section{RESULTS AND DISCUSSION}\label{sec:result}

In this section, we turn to the presentation of our numerical results for electronic structure with SOC, phonon dispersion relation, EPC by making use of the scattering time and Eliashberg spectral function, the current-current response tensor and optical absorption of the system described in preceding section. Finally, we discuss the charge collective mode of the system in the presence of the SOC and EPC in the long-wavelength limit.

\subsection{Electron and phonon band structures of antimonene}
Antimonene has a buckled honeycomb lattice with an indirect band gap as illustrated in Figs.\,\ref{fig1}(a) and \ref{fig1}(b), respectively. The calculated relaxed lattice parameter of antimonene is found to be $a=4.0\,{\rm \AA}$  and two sublattices are vertically displaced by $b=1.61\,{\rm \AA}$. The indirect band gap is calculated to be $1.05$\,eV, the valence band maximum (VBM) and conduction band minimum (CBM) are located at $\Gamma$ point and between line $\Gamma-M$, respectively. This value is in good agreement with the previous DFT-LDA result\,\cite{rudenko2017}. By including the SOC, we observe a considerable reduction of the band gap to about $0.74$\,eV. As a result, it is essential to include the SOC contribution in our calculations. It is worth mentioning that the electronic band structure we obtained is similar to those results calculated by different groups only with the energy band gap value smaller than the value they calculated for a pristine antimonene~\cite{xu2017first, zhang2017antimonene,chen2016electronic}.  It turns out that the energy band gap of antimonene depends strongly on the exchange-correlation term and basis which are implemented into DFT codes.

The small band gap of less than 1.2 eV indicates that antimonene will not be optically transparent, as there will be an appreciable amount of free carriers created by absorption even at room temperature. Therefore, antimonene will start to absorb electromagnetic radiation in the infrared region as an optical material.

The electronic band structure of antimonene along the high-symmetry points $\Gamma$-M-K-$\Gamma$ of the Brillouin zone is shown in Fig.\,\ref{fig1}(b). The maximum of the valence band is found at the $\Gamma$ point, while the conduction band minimum is between $\Gamma$ and M points. Furthermore, it clearly shows a high degree of electron-hole asymmetry even for very low doping concentrations. These results also show the impact of the SOC on the electronic band structure. Namely, the SOC lifts the degeneracy of the two topmost valence bands at $\Gamma$ point, as well as reduces the band gap by about 0.3\,eV.

The phonon spectrum of antimonene, on the other hand, is plotted in Fig.\,\ref{fig1}(c).
The in-plane acoustic modes
display a linear dispersion near the zone center, whereas the out of plane, ZA, branch has a parabolic dispersion in a similar manner to that of graphene.
In contrast with the electronic structure, the SOC has a negligible impact on the phonon spectra. However, the impact of the SOC is relatively stronger for optical phonon modes and it is less prominent for acoustic modes\,\cite{Lugovskoi2019}.

It is technically useful to consider a linear behavior of the dispersion relation in the long-wavelength limit to obtain the phonon sound velocities. We thus calculate the slopes of in-plane acoustic branches in the vicinity of the $\Gamma$ point.
The derived sound velocities are 
about $3.60, 2.4$ km/s for longitudinal and transverse atomic motions, respectively, along the $\Gamma-K$
direction. Along the
$\Gamma-M$ direction, on the other hand, the sound velocities are obtained as $3.61$ and $2.31$ km/s for longitudinal and transverse atomic motions, respectively. Therefore, a weak anisotropy within the entire
first Brillouin zone is predicted for sound velocities in monolayer antimonene.  It
is worth mentioning that these sound velocities are comparable to
those obtained for other 2D materials, such as graphene, $14.9 - 21.8$
 km/s \cite{zou2016phonon}, MoS$_2$,  $4.2 - 6.8
$ km/s \cite{liu2014anisotropic}, monolayer phosphorene, $4.48-7.59$ km/s~\cite{PhysRevB.91.115412}, blue phosphorene, $4.0 - 8.0
$ km/s~\cite{esfahani2017superconducting, jain2015strongly} and stanene,
$1.3 - 3.6 $ km/s \cite{peng2016low}.

\subsection{Optical absorption}
We explore the optical absorption calculated by using Eqs.~(\ref{intra_cu}-\ref{landa}) and making use of the approximation in which $q$ tends to zero. Having calculated the electronic structure with SOC, we can obtain the Kohn-Sham wave functions and energies, which are invoked to calculate the EPC value and the current-current response tensor. 
Basically, we solve the screened current-current response tensor in the quasi-2D crystal utilizing $\Pi=\Pi^0 + \Pi^0 \otimes D^0 \otimes \Pi$.
In this approach, the momenta dependence actually originates from the free-photon propagator $D^0(q,\omega)$. Furthermore, we only consider a normal incident light with $s$ polarization.

\begin{figure}
\begin{center}
\includegraphics*[width=6.4cm]{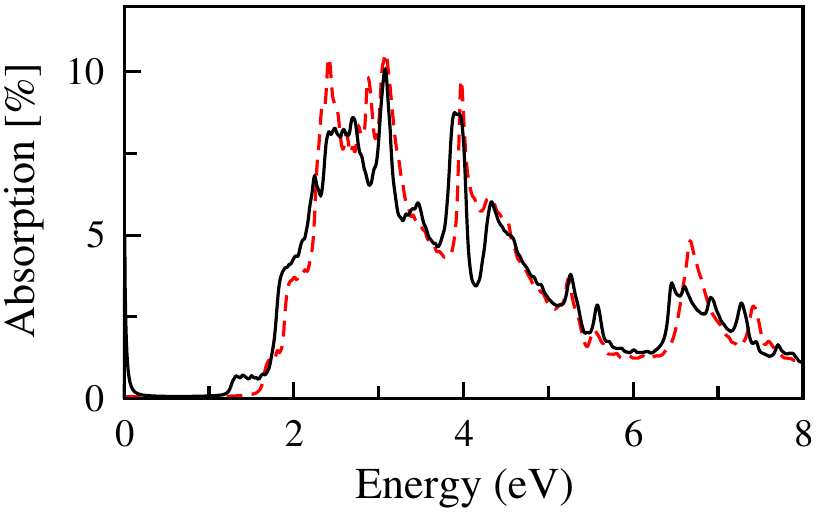}
\caption{(Color online) The optical absorption of monolayer $\beta$-Sb as a function of energy with (black line) and without (red dashed-line) spin-orbit coupling at $q=0$ for normal incidence. Wide optical absorption in the energy range of $2-4$ eV occurs. The intensive peaks of the optical absorption are at 2.4 and 3.1\,eV. By including the SOC, the intensive peak at 2.4\,eV splits and its intensity decreases.}\label{fig2}
\end{center}
\end{figure}

The optical absorption of antimonene, which is based on the many-body processes, with (black line) and without (red dashed-line) SOC is shown in Fig.\,\ref{fig2}. The excitation energy corresponding to the first absorption peak which mainly originates from the interband transition between the VBM and CBM. Stemming from our DFT analysis, the orbital character of the conduction and valence bands are $0.3 |p_x\rangle+0.3 |p_y\rangle+0.4|s\rangle$ and $0.5|p_x\rangle+0.5|p_y\rangle$, respectively. Accordingly, the optical transition from VBM to CBM is allowed.

Most importantly, there is an extensive optical absorption in the energy range of $2-4$ eV (including the main part of the visible light and ultraviolet), which makes the $\beta$-Sb as a potential
candidate material for photovoltaic devices\,\cite{shu2018}.
The intensive peaks of the optical absorption are at around 2.4 and 3.1\,eV. By including the SOC, the intensive peak at 2.4\,eV splits and its intensity decrease. Moreover, the optical absorption onset starts at lower energy, i.e., around 1.18\,eV, compared with the optical absorption without SOC,  which is a result of the SOC-induced reduction of the band gap.

\subsection{Rigid band approximation of carrier doping}
For the sake of simplicity, we first consider a rigid band model which is a good starting point to describe the doping effects, since the corresponding band structure changes are small. In this approach, the electronic band structure and lattice dynamics are assumed to be unaffected by the presence of doped electrons and holes. The degree of doping is determined by the placement of the Fermi energy into the conduction and valence bands for electron and hole doping, respectively. 
\par
In order to perceive the effect of the electron and hole doping over a wide but experimentally accessible range, we consider $0.1$ and $0.2$ holes per unit cell, which corresponds to carrier concentrations of $7.3\times 10^{13}$\,cm$^{-2}$ and $1.4\times 10^{14}$\,cm$^{-2}$. In order to reach these concentrations, the Fermi energy should be set to $0.55$ (0.35) and $0.77$ (0.61) eV below the VBM for SOC (non-SOC) calculations. In the electron-doped case, we consider $0.1$ electrons per unit cell that corresponds to the Fermi energy of about $0.13$ and $0.12$\,eV above the CBM for SOC and non-SOC calculations, respectively. The black dotted lines in Fig.~\ref{fig1}(b) shows the Fermi energy of different dopings for SOC calculations.  

Having used the previous approach, we would like to study the charge plasmon mode in the system. Electron energy-loss spectroscopy (EELS) is an analytical technique which is based on inelastic scattering of fast electrons in a thin sample and is defined as
\begin{equation}
L(q,\omega)=-{\rm Im} (\frac{1}{\varepsilon(q,\omega)})
\end{equation}
where the many-body dielectric function is defined as $\frac{1}{\varepsilon(q,\omega)}=1+v_q \chi_{nn}(q,\omega)$ in which $\chi_{nn}(q,\omega)$ is the charge-charge response function of the system and $v_q$ is the bare Coulomb potential. By making use of the relation between the charge-charge and longitudinal current-current response functions for a $s$ polarization together with the fluctuation-dissipation relation, we end up to a relation in which $L(q,\omega)=\pi e^2 A(q,\omega)$. This relation tells us that the spectrum of the peak of $A(q,\omega)$ can be considered as a plasmon modes of the system.

\begin{figure}
\begin{center}
\includegraphics*[width=8.4cm]{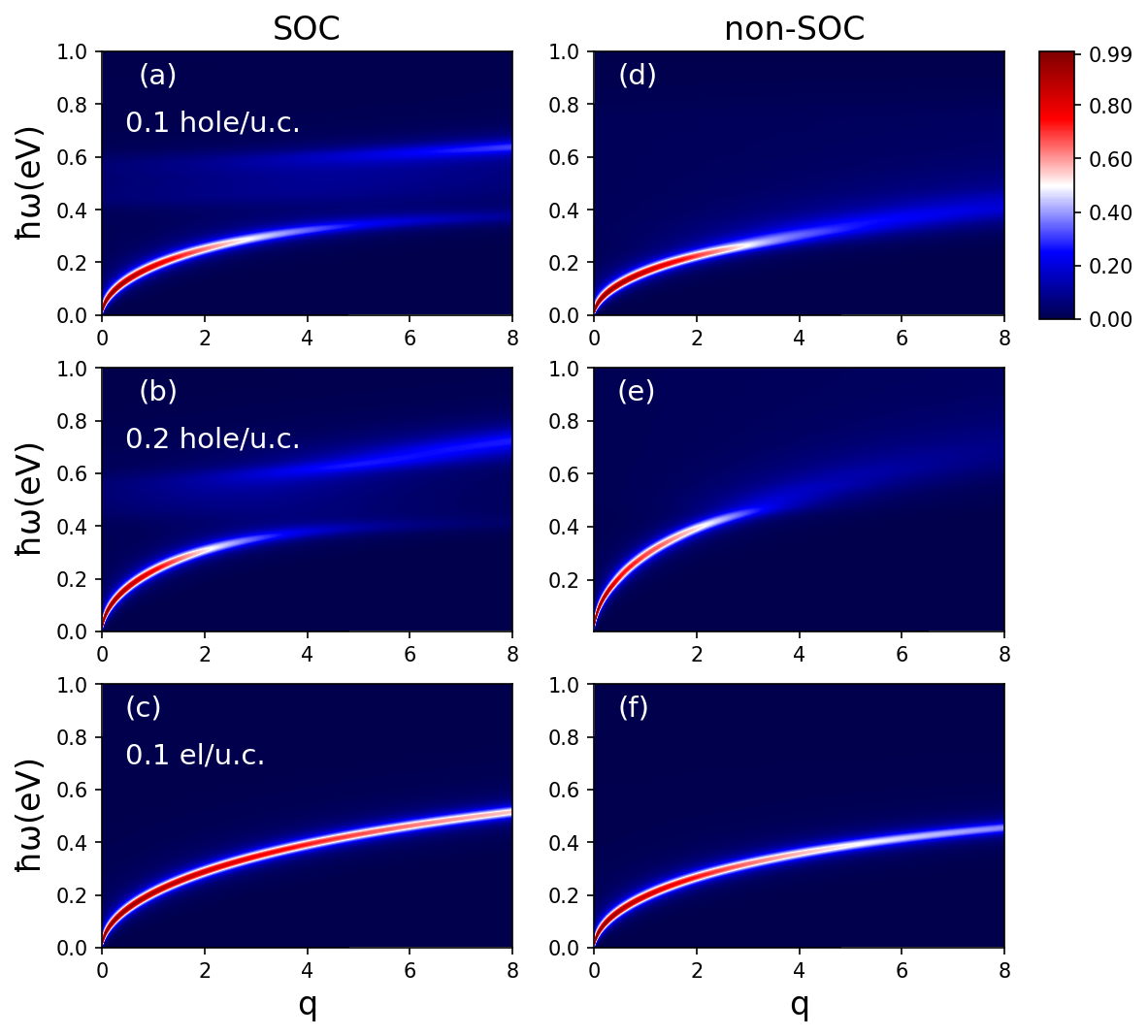}
\caption{(Color online) Charge plasmon modes as a function of $q$ (in units of $10^{-3}$ bohr$^{-1}$) within the rigid band approximation with SOC (a)-(c) and without SOC (d)-(f) for $0.1$ and $0.2$ holes/u.c. and $0.1$ electrons/u.c. In the hole doping case, the splitting of plasmon branches is occurred for the SOC, while it is absent in the case of without SOC. Basically, the modified electronic band structure at the $\Gamma$ point in the valence band owing to the SOC, make manifest itself explicitly in the plasmon dispersion. }\label{fig3}
\end{center}
\end{figure}

Fig.\,\ref{fig3} displays the plasmon dispersion of doped antimonene in which (a)-(c) denote results with SOC, while (d)-(f) refer to results without SOC. The main difference between the SOC and non-SOC cases for the hole doping is the splitting of plasmon branches that occurs for the former, while it is absent for the latter. Also, the second branches of plasmon modes become wider by increasing the hole concentration. In fact, the SOC term causes a reconstruction of the electronic band structure and lifts the degeneracy of two bands at the $\Gamma$ point in the valence band, which manifests itself explicitly in the plasmon dispersion. For 0.1 electrons/u.c. doping, on the other hand, there is no noticeable difference for the plasmon dispersions with and without SOC term. 

\begin{figure}
\begin{center}
\includegraphics*[width=8.6cm]{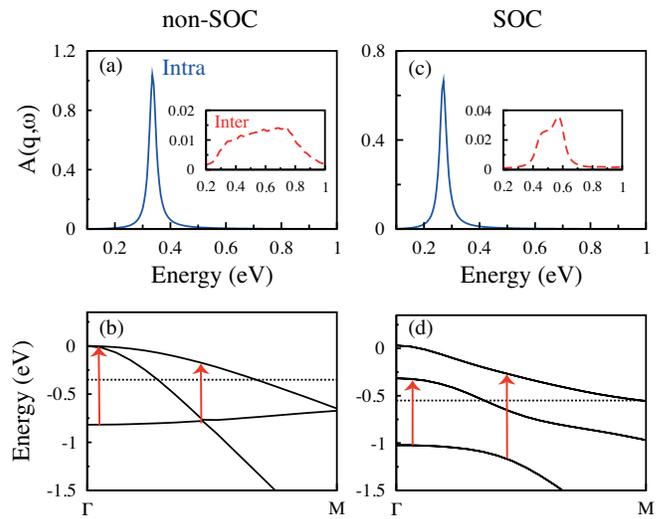}
\caption{(Color online) The intra- and interband contributions of the excitation spectra $A(q,\omega)$ at $q= 0.002$ bohr$^{-1}$ with the charge concentration of $0.1$ holes/u.c. within the rigid band approximation. (a) and (c) demonstrate $A(q,\omega)$ without and with SOC, respectively. (b) and (d) illustrate the close-up of the band structure around the band splitting at $\Gamma$ point for SOC and non-SOC cases. The arrows represent that interband transitions occur. The Fermi energy is shown by black dotted lines.}\label{fig4}
\end{center}
\end{figure}

For the sake of completeness, we plot the intra- and interband excitation spectra as function of energy at $q=0.002$ bohr$^{-1}$ with the charge concentration of $0.1$ hole/u.c. as well as with and without SOC in Fig.~\ref{fig4}(a) and \ref{fig4}(c). The results show that the second branch of the plasmon dispersion given in Figs.~\ref{fig3}(a) and \ref{fig3}(b) orginates from the low-energy interband transitions and it can be related to the SOC-induced band splitting. These interband transitions come from the topmost three valence bands somewhere around the $\Gamma$ point. Figs.~\ref{fig4}(b) and \ref{fig4}(d) show the close-up of the band structure around the band splitting at $\Gamma$ point for 0.1 holes per unit cell doping for SOC and non-SOC cases. The arrows represent approximately the relevant interband transitions.

\par 

\subsection{Carrier doping within jellium model}
In the following, we study plasmon dispersion in antimonene when the electron and hole dopings are obtained within a more realistic model, i.e., a jellium model. 
In the jellium model, carrier doping is simulated by adding or removing excess electronic charge into the unit cell which is compensated by a uniform positive background.

For every doping case, we calculate the plasmon spectrum of antimonene for a fully optimized lattice parameter (variable lattice parameter), keeping the same vacuum space, as well as when the lattice parameter is fixed. The fixed lattice parameter is set to the lattice parameter of the pristine antimonene. The latter could be regarded as a case that models the behavior of the 2D material adsorbed on a substrate\,\cite{Lugovskoi2019}. The lattice parameters $a$, bond length $l$, buckling height $\Delta$ and bond angle $\theta$ for the various doping levels of variable and fixed lattice parameters are summarized in Table\,\ref{tab:table1}. For the variable lattice parameter, the antimonene has a tendency to decrease its lattice parameter and bond angle in the hole-doped case, while the bond angle increases for electron doping. The lattice parameter can be affected less by electron doping. But buckling height has opposite behavior, it increases in the hole-doped case and decreases for the electron-doped case.  Also, the bond length is less sensitive to doping compared to the buckling height and bond angle. The similar behavior is observed in the case of the fixed lattice parameter. 

\begin{table}[!t]
\caption{\label{tab:table1}
Structural parameter of a doped antimonene. Lattice parameter for variable lattice parameter $a$,  bond length
 $l$ and buckling height $\Delta$ (\AA), together with the bond angle $\theta$ ($^{\circ}$) for variable and fix lattice parameters.}
\begin{center}
\begin{tabular}{cccccccc}
\hline
\hline
  &    &\multicolumn{3}{c}{variable lattice}&\multicolumn{3}{c}{fixed lattice}\\\cline{3-5}\cline{6-8}
     doping   &   $a$    &  $l$ & $\Delta$ & $\theta$ & $l$   & $\Delta$ & $\theta$   \\ 
\hline
0.1 holes/u.c. & 3.906 & 2.819 & 1.691 & 87.71 & 2.839  & 1.661 & 89.21 \\ 
0.2 holes/u.c. & 3.840 & 2.828 & 1.756 & 85.50 & 2.861 & 1.700 & 88.32\\
\hline
~0.1 electrons/u.c.~~& 4.020  & 2.823 & 1.606 & 90.81& 2.815 & 1.620 & 90.18\\
\hline
pristine~~~~  & 4.000   & 2.817 & 1.623 & 90.10&        &       &    \\  
\hline
\hline
\end{tabular}
\end{center}
\end{table}

\begin{figure}
\begin{center}
\includegraphics*[width=8.4cm]{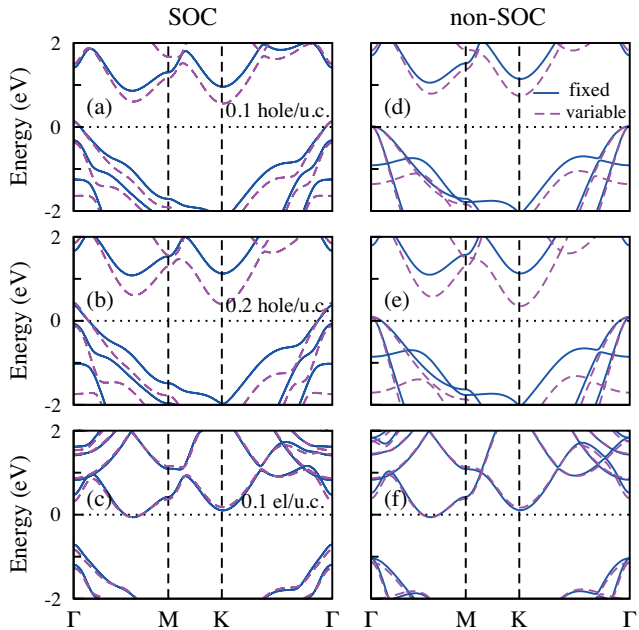}
\caption{(Color online) The electronic band structure for $0.1$ and $0.2$ holes/u.c. and $0.1$ electrons/u.c. The band structure of the variable unit cell and the fixed unit cell when the carrier doping is performed by means of the jellium model are shown by a magenta dashed line and blue line, respectively. (a)-(c) The results by including SOC and (d)-(f) non-SOC ones. The Fermi energy is set to be zero.}\label{fig5}
\end{center}
\end{figure}

It is imperative to understand how the electronic and plasmon dispersion can be modulated at different doping levels.
Fig. \ref{fig5} displays the evolution of the electronic band structure upon a hole and electron doping for both the variable lattice and fixed lattice parameters with and without including spin-orbit interaction. In the case of hole doping (both with and without the SOC) the valence bands are almost unchanged, especially near the Fermi energy. For the variable lattice method, on the other hand, the lowest energy conduction band drastically changes and it shifts downward when the concentration of holes is elevated. At the same time, the CBM edge is slightly shifted from the middle of the $\Gamma$-$\it M$ pathway toward the $\it M$ point. On the other hand, the band structure changes are absent for the fixed lattice parameter. In contrast to the hole- doped, there is no remarkable difference in the electron-doped case. 

\begin{figure}
\begin{center}
\includegraphics*[width=8.4cm]{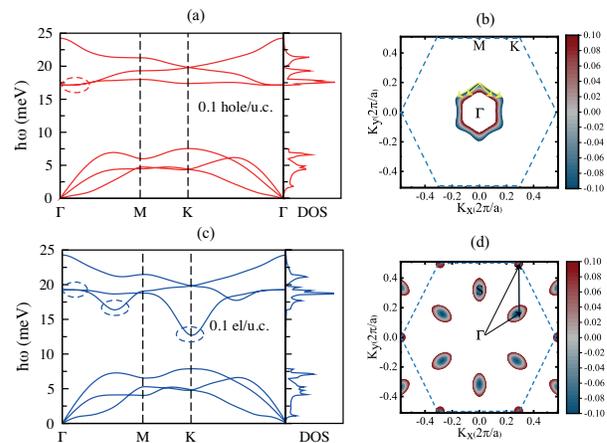}
\caption{(Color online) The phonon spectra and Fermi surface topology of the monolayer antimonene when the doping is simulated within the jellium model. (a) and (c) demonstrate phonon dispersion and related density of states (DOS) of 0.1 holes/u.c. and 0.1 electrons/u.c. doped concentration. The sketch of the Fermi surface topology is shown in (b) and (d) for  0.1 holes/u.c. and 0.1 electrons/u.c. over the first Brillouin zone, respectively.  The specific phonon wave vectors lead to the Kohn anomaly are represented by dashed elliptical lines in the left-hand graphs and solid yellow (black) lines in the right-hand graphs referring to specific phonon wave vectors used in intra(inter) layer scattering process.}
\label{fig6}  
\end{center}
\end{figure}

The calculated phonon spectra and phonon DOS of 0.1\,electrons/u.c. and 0.1\,holes/u.c. doping for antimonene are given in Fig.\,\ref{fig6}. It is worth mentioning that Kohn anomalies\,\cite{kohn1959image,lazzeri2006nonadiabatic} (the itinerant charge carriers of the doped antimonene renormalize the bare phonon frequencies and cause an anomaly in the phonon dispersion) appear both in the electron- and hole-doped cases and are shown by blue and red ellipses, respectively, in the figure. In the electron doped case, the Kohn anomalies appear for the optical phonon at the $\it K$ point as well as along the $\Gamma$-$\it M$ (which we indicate here as the S point). The phonon softening originates from the relatively strong adiabatic electron-phonon renormalization (i.e., EPC at $q={\rm K}$ and $q={\rm S}$).  
In the case of hole doped antimonene, the Kohn anomaly appears at the $q=\Gamma$ point, since in this case $k=\Gamma$ is partially depopulated and the corresponding Fermi surface plots are shown in Figs.\,\ref{fig6}(b) and \,\ref{fig6}(d).

\begin{figure}
\begin{center}
\includegraphics*[width=8.4cm]{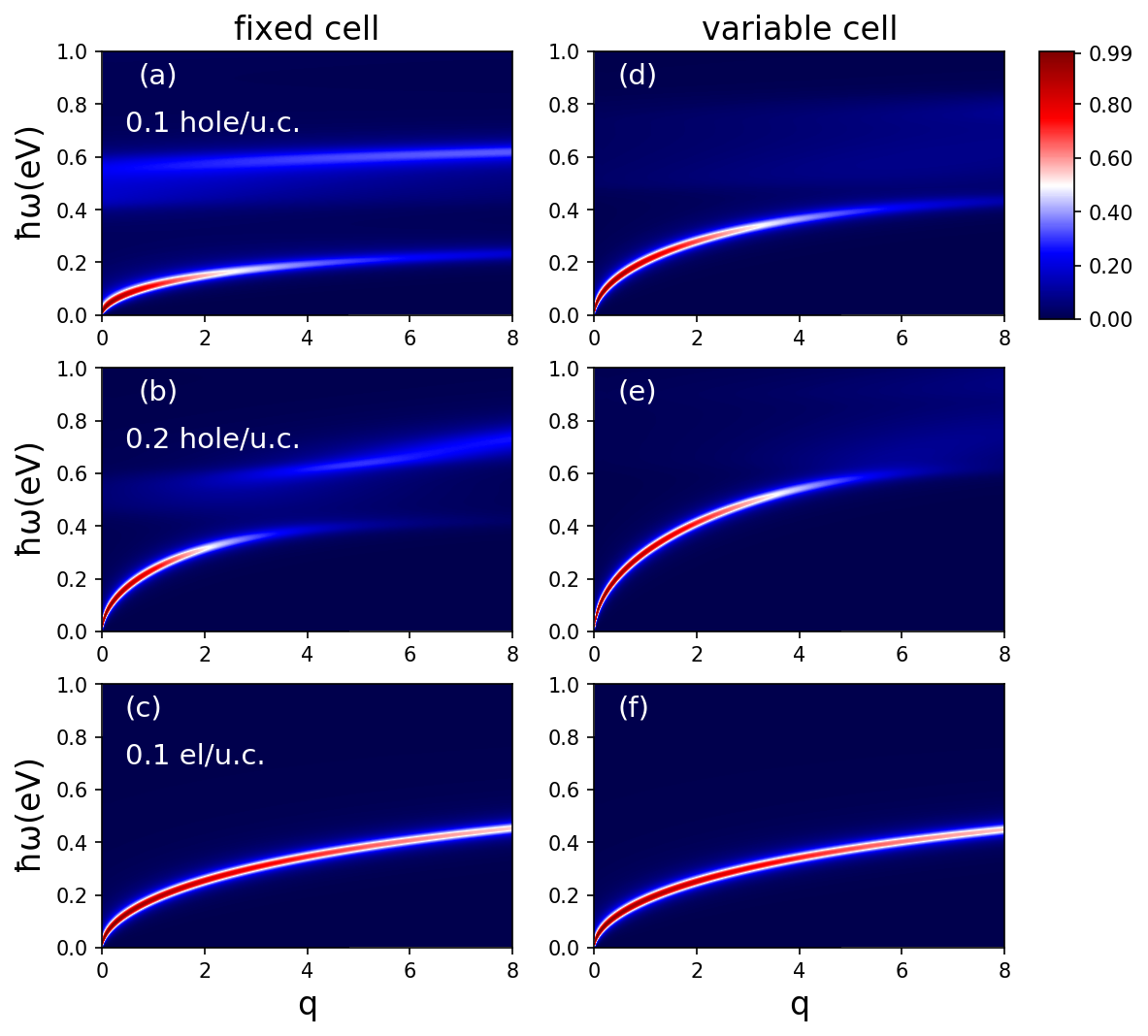}
\caption{(Color online) The charge plasmon mode as a function of $q$ (in units of $10^{-3}$ bohr$^{-1}$) of antimonene for the same carrier concentrations as in Fig.\,\ref{fig4}, however, carried on within the jellium model. (a)-(c) are the results for the fixed unit cell while (d)-(f) for the variable unit cell. The SOC is considered for all of calculations. Notice that the plasmon dispersions of variable lattice parameters are significantly different from the ones when the lattice parameter is kept fixed.}\label{fig7}
\end{center}
\end{figure}

In Fig.\,\ref{fig7}, we compare the plasmon dispersion of antimonene for both the variable and fixed lattice parameters. All calculations undertaken in this figure are with SOC. It can be seen that for the hole doping, the plasmon dispersions of variable lattice parameters are completely different from the ones when the lattice parameter is kept fixed. The splitting of plasmon dispersion for both the 0.1 and 0.2\,holes/u.c. disappear when fully relaxed calculations are used. It is also interesting that the plasmon dispersions with the fixed lattice parameter [i.e., Fig.\,\ref{fig7}(a) and \ref{fig7}(b)] are similar to the results of plasmon modes within the rigid band approximation presented in Fig.\,\ref{fig3}(a) and \ref{fig3}(b).

In the electron doped case, there is no notable difference for both the variable lattice and fixed lattice parameter, because of no considerable change in the corresponding band structures shown in Fig.\,\ref{fig5}(c) and \ref{fig5}(f).


 
\subsection{Phonon-induced plasmon decay in antimonene}
\begin{figure}
\begin{center}
\includegraphics*[width=10.4cm]{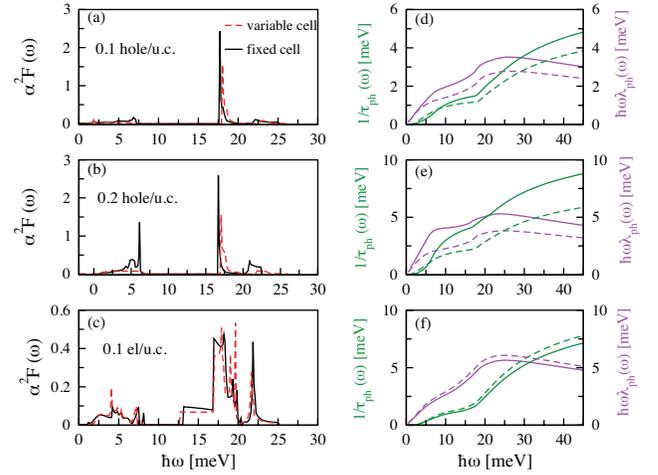}
\caption{(Color online) (a)-(c) Calculated Eliashberg spectra, (d)-(f) the phonon-induced decay rate $1/\tau_{\rm ph}(\omega)$ and renormalization parameter $\omega\lambda_{\rm ph}(\omega)$ for $0.1$ and $0.2$ holes/u.c. and $0.1$ electrons/u.c. for monolayer Sb$_2$ for the fixed unit cell (solid lines) within the jellium model. The corresponding results with the variable unit cell are shown by dashed lines.}\label{fig8}
\end{center}
\end{figure}

Now, we would like to investigate the impact of the EPC on the plasmon dispersion. For this purpose, we consider electron- and hole-doped cases. By making use of the calculated Eliashberg function, $\alpha^2F(\omega)$, and Eqs.\,(\ref{tau}) and (\ref{landa}), we can obtain the phonon-induced decay rate $1/\tau_{\rm ph}(\omega)$ and renormalization parameter $\omega\lambda_{\rm ph}(\omega)$ for electron and hole doped cases (see Fig.\,\ref{fig8}). Notice, $\alpha^2F(\omega)$ is calculated on $24\times24\times1$ $\textbf k$-points and $8\times8\times1$
$\textbf q$-mesh. While, a finer $\textbf k$-mesh $120\times120\times1$ is applied for calculating the EPC.

\begin{figure}
\begin{center}
\includegraphics*[width=8.4cm]{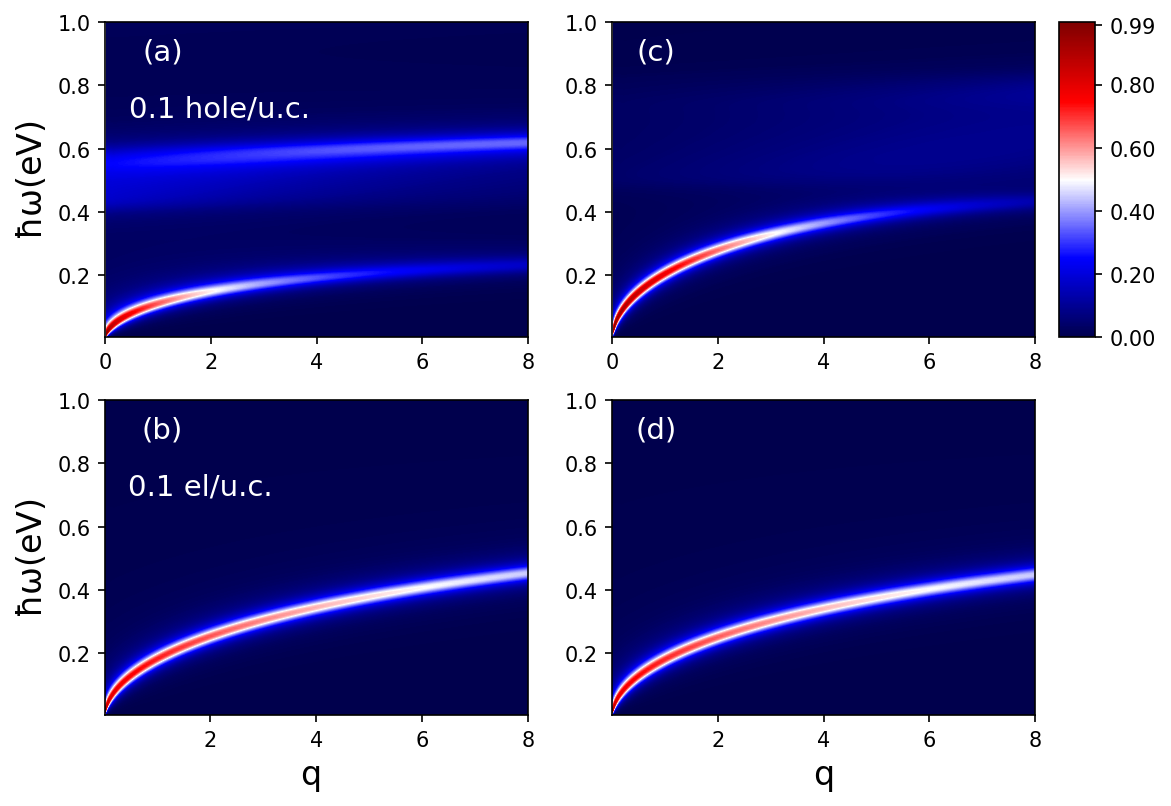}
\caption{(Color online) Plasmon dispersion as a function of $q$ (in units of $10^{-3}$ bohr$^{-1}$) for $0.1$ holes/u.c. and $0.1$ electrons/u.c. for monolayer Sb$_2$ with taking into account the electron-phonon coupling. (a) and (b) are the results for the fixed unit cell while (c) and (d) for the variable unit cell. The SOC contribution is considered in all calculations.}\label{fig9}
\end{center}
\end{figure}

\begin{figure}
\begin{center}
\includegraphics*[width=8.4cm]{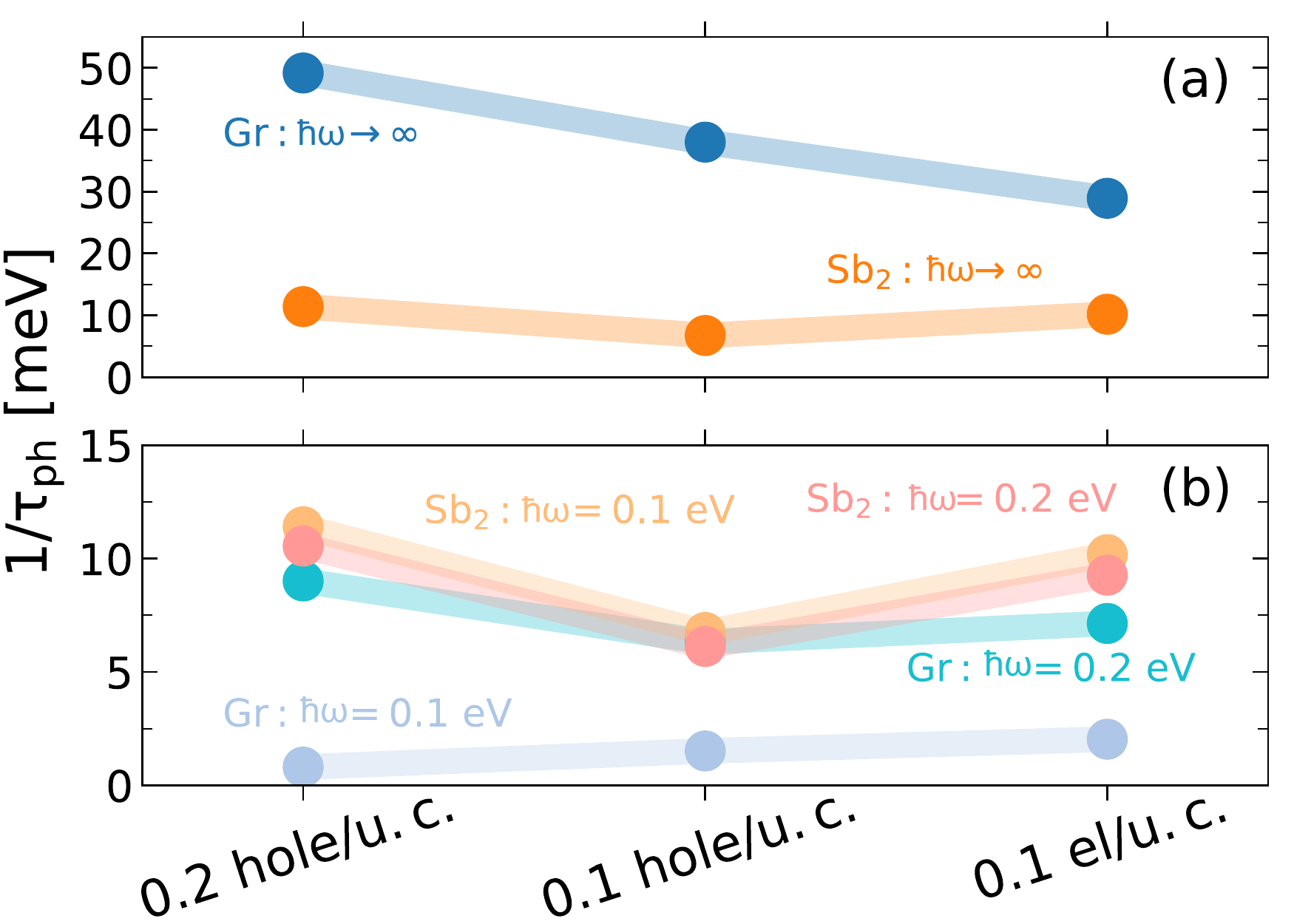}
\caption{(Color online) The plasmon damping rates due to the electron-phonon coupling $1/\tau_{\rm ph}(\omega)$ (in units of $\hbar$) of monolayer graphene and antimonene compared for $\hbar \omega= 0.1$\,eV,  $\hbar \omega= 0.2$\,eV and in the high-energy limit ($\hbar \omega \rightarrow \infty$) for 0.1\,holes/u.c., 0.2\,holes/u.c. and 0.1\,electrons/u.c. In panel (a) the high-energy limit results of $1/\tau_{\rm ph}( \omega)$, while in (b) results of $\hbar \omega= 0.1$\,eV and 0.2\,eV are compared.}\label{fig10}
\end{center}
\end{figure}

It can be seen that the same behavior is obtained from results calculated from the fixed unit cell and the variable lattice parameter approaches. The main discrepancy between the two cases is that the Eliashberg function of the fixed unit cell has a slightly more (less) intensity in comparison with the variable lattice parameter for hole-doped (electron-doped) and it increases by growing the doping level. 

From a more detailed inspection, it is clear that the largest peak in $\alpha^2F(\omega)$ for 0.1 holes/u.c. stems from the Kohn anomaly owing to the EPC of the optical in-plane displacements at $q \approx \Gamma$. The existing hole pocket around $k=\Gamma$ point reveals that only intralayer scatterings from one point of the Fermi surface to another one is feasible for this doping level [see yellow arrows in Fig.\,\ref{fig6}(b)]. The latter scattering processes are thus responsible for the Kohn anomaly and for the plasmon damping in the hole-doped case. For larger hole concentrations (0.2 holes/u.c.), the form of the Fermi surface is generally similar to a lower doping level. Therefore, the $\alpha^2F(\omega)$ is not modified a lot when the hole concentration increases due to the lack of the change in the type and the number of hole pockets. Nevertheless, there are still some small differences in the spectral function for the two cases that are attributed to a slight increase of the hole pocket existing at the vicinity of the $\Gamma$ point, but also to appearance of a new coaxial pocket at the around the $\Gamma$ point and very close to the previous pockets, which originates from the SOC-induced bands splitting.

It should be pointed out that for the electron-doped case an additional spectral feature appears in $\alpha^2F(\omega)$, i.e., a nonzero plateau around the phonon energy of 15\,meV. Such a feature arises from the existence of the strong Kohn anomalies at $q={\rm K}$ and $q={\rm S}$ points, as depicted in the Fig.\,\ref{fig6}(c) by dashed elliptical lines. These Kohn anomalies and strong EPC at these points, and thus the aforesaid plateau in $\alpha^2F(\omega)$, are prompted by a suitable form of the Fermi surface [see Fig.\,\ref{fig6}(d)]. Namely, these spectral features visible in $\alpha^2F(\omega)$ are due to interlayer scattering between $k={\rm K}$ and $k={\rm S}$ valleys assisted with the phonon wavevectors $q={\rm S}$ as well as between inequivalent $k={\rm S}$ valleys assisted with $q={\rm K}$ phonon, as depicted in Fig.\,\ref{fig6}(d) by black arrows. Note that intralayer scatterings within both $k={\rm K}$ and $k={\rm S}$ electron pockets are present in the electron doped case.

The plasmon dispersion of electron- and hole-doped antimonene is calculated by the approach introduced in Sec.\,\ref{method}. The results are exhibited in Fig.\,\ref{fig9} for $0.1$ holes/u.c. and $0.1$ electrons/u.c. with taking into account the EPC. It is completely explicit, the EPC increases
 the peak width of $A(q,\omega)$.
It is important to mention that owing to a small EPC the phonon-induced plasmon decay of antimonene is quite small.
As aforementioned discussion points out, the plasmons in antimonene decay mostly due to phonon scatterings around $q=\Gamma$ point (intravalley transitions) for the hole doped case, while the damping in the electron doped case is a bit complexed, where both intravalley ($q={\rm \Gamma}$) and intervalley ($q={\rm K}$ and $q={\rm S}$) scatterings take place.

In order to emphasize the low-loss plasmonic properties in antimonene, the corresponding values of scattering rates due to EPC, $1/\tau_{\rm ph}(\hbar \omega)$ for three values of $\hbar \omega$ (i.e., $\hbar \omega=0.1$\,eV, 0.2\,eV and for the high-energy limit $\omega \rightarrow \infty$) and for total charges of 0.1\,holes/u.c. and 0.2\,holes/u.c., and 0.1\,electrons/u.c. are compared with the scattering rates in well-known graphene [see Figs.\,\ref{fig10}(a) and \,\ref{fig10}(b)]. Note that $1/\tau_{\rm ph}(\infty) = \pi \sum_{{\bf q} \nu} \omega_{{\bf q} \nu}  \lambda_{{\bf q} \nu}$, where $\lambda_{{\bf q} \nu}$ is the standard electron-phonon coupling constant and the summation goes over all ${\bf q}$ and phonon bands\,\cite{novko17}. While $1/\tau_{\rm ph}$ in the case of antimonene is more or less the same for all three values of $\omega$ (since $\omega$ is always larger than the characteristic phonon energies), for graphene it changes. These three values are important for graphene, since $1/\tau_{\rm ph}(\hbar \omega=0.1\,{\rm eV})$ includes mostly contributions from acoustic modes, $1/\tau_{\rm ph}(\hbar \omega=0.2\,{\rm eV})$ includes mostly contributions from acoustic and some of the optical modes, while $1/\tau_{\rm ph}(\omega=\infty)$ includes all possible modes\,\cite{principi14,novko17}. It can be seen that when only acoustic and some of the optical modes are active in graphene (i.e., when plasmon energy $\hbar \omega\lesssim 0.2$\,eV) plasmon decay rates are comparable in graphene and antimonene [see Figs.\ref{fig10}(b)]. On the other hand, for larger plasmon energies ($\hbar \omega> 0.2$\,eV) antimonene retains its low-loss properties, while plasmon decay rate in graphene is significantly elevated [see Figs.\,\ref{fig10}(a)]. In other words, plasmons in antimonene are significantly less damped compared to graphene when plasmon energies are $\hbar \omega> 0.2$\,eV due to smaller plasmon-phonon coupling in the former material.



%
\section{CONCLUSION}\label{sec:concl}
In summary, we have investigated the electronic and optical properties of electron- and hole-doped $\beta$-phase antimonene utilizing first-principles calculations. We have also considered the effect of charge carrier doping effects by making use of a realistic approach based on the jellium model. 

Our results show that antimonene possesses a very interesting electron excitation spectra. In particular, plasmon modes are quite sensitive to the SOC and are characterized by very low damping rates due to small electron-phonon scatterings. The plasmons in antimonene decay mostly due to phonon scatterings around $q=\Gamma$ point (intravalley transitions) for the hole doped case, while the damping in the electron doped case, where both intravalley ($q={\rm \Gamma}$) and intervalley ($q={\rm K}$ and $q={\rm S}$) scatterings. Furthermore, the scattering rates owing to the EPC of antimonene are compared with the scattering rates in monolayer graphene. We have concluded that when only acoustic and some of the optical phonon modes are active in graphene (i.e., when plasmon energy $\hbar \omega\lesssim 0.2$\,eV) plasmon decay rates in antimonene are comparable in those obtained in graphene, however, antimonene retains its low-loss properties for larger plasmon energies ($\hbar \omega> 0.2$\,eV) where plasmon decay rate in graphene is significantly elevated. Our finding suggests that this novel 2D material is a promising material with potential applications in next-generation plasmonic devices.

\begin{acknowledgments}
This work is supported by the Iran Science Elites Federation.
D.N. acknowledges financial support from the Croatian Science Foundation (Grant no. UIP-2019-04-6869) and from the European Regional Development Fund for the ``Center of Excellence for Advanced Materials and Sensing Devices'' (Grant No. KK.01.1.1.01.0001).
\end{acknowledgments}

\bibliographystyle{unsrt}
\bibliography{sb2}

\begin{thebibliography}{10}

\bibitem{novoselov2004electric}
Kostya~S Novoselov, Andre~K Geim, Sergei~V Morozov, D~Jiang, Y\_ Zhang,
  Sergey~V Dubonos, Irina~V Grigorieva, and Alexandr~A Firsov.
\newblock Electric field effect in atomically thin carbon films.
\newblock {\em science}, 306(5696):666--669, 2004.

\bibitem{geim2013van}
Andre~K Geim and Irina~V Grigorieva.
\newblock Van der waals heterostructures.
\newblock {\em Nature}, 499(7459):419--425, 2013.

\bibitem{mounet2018two}
Nicolas Mounet, Marco Gibertini, Philippe Schwaller, Davide Campi, Andrius
  Merkys, Antimo Marrazzo, Thibault Sohier, Ivano~Eligio Castelli, Andrea
  Cepellotti, Giovanni Pizzi, et~al.
\newblock Two-dimensional materials from high-throughput computational
  exfoliation of experimentally known compounds.
\newblock {\em Nature nanotechnology}, 13(3):246--252, 2018.

\bibitem{stability}
Pablo Ares, Fernando Aguilar-Galindo, David Rodr{\'\i}guez-San-Miguel, Diego~A
  Aldave, Sergio D{\'\i}az-Tendero, Manuel Alcam{\'\i}, Fernando Mart{\'\i}n,
  Julio G{\'o}mez-Herrero, and F{\'e}lix Zamora.
\newblock Mechanical isolation of highly stable antimonene under ambient
  conditions.
\newblock {\em Advanced Materials}, 28(30):6332--6336, 2016.

\bibitem{ares2018recent}
Pablo Ares, Juan~Jos{\'e} Palacios, Gonzalo Abell{\'a}n, Julio
  G{\'o}mez-Herrero, and F{\'e}lix Zamora.
\newblock Recent progress on antimonene: a new bidimensional material.
\newblock {\em Advanced Materials}, 30(2):1703771, 2018.

\bibitem{ji2016two}
Jianping Ji, Xiufeng Song, Jizi Liu, Zhong Yan, Chengxue Huo, Shengli Zhang,
  Meng Su, Lei Liao, Wenhui Wang, Zhenhua Ni, et~al.
\newblock Two-dimensional antimonene single crystals grown by van der waals
  epitaxy.
\newblock {\em Nature communications}, 7(1):1--9, 2016.

\bibitem{wu2017epitaxial}
Xu~Wu, Yan Shao, Hang Liu, Zili Feng, Ye-Liang Wang, Jia-Tao Sun, Chen Liu,
  Jia-Ou Wang, Zhong-Liu Liu, Shi-Yu Zhu, et~al.
\newblock Epitaxial growth and air-stability of monolayer antimonene on pdte2.
\newblock {\em Advanced Materials}, 29(11):1605407, 2017.

\bibitem{shi2019van}
Zhi-Qiang Shi, Huiping Li, Qian-Qian Yuan, Ye-Heng Song, Yang-Yang Lv, Wei Shi,
  Zhen-Yu Jia, Libo Gao, Yan-Bin Chen, Wenguang Zhu, et~al.
\newblock van der waals heteroepitaxial growth of monolayer sb in a puckered
  honeycomb structure.
\newblock {\em Advanced Materials}, 31(5):1806130, 2019.

\bibitem{high_mobility}
Long Cheng, Chenmu Zhang, and Yuanyue Liu.
\newblock The optimal electronic structure for high-mobility 2d semiconductors:
  Exceptionally high hole mobility in 2d antimony.
\newblock {\em Journal of the American Chemical Society}, 141(41):16296--16302,
  2019.

\bibitem{kurpas2019spin}
Marcin Kurpas, Paulo E~Faria Junior, Martin Gmitra, and Jaroslav Fabian.
\newblock Spin-orbit coupling in elemental two-dimensional materials.
\newblock {\em Physical Review B}, 100(12):125422, 2019.

\bibitem{rudenko2017}
AN~Rudenko, MI~Katsnelson, and Rafael Rold{\'a}n.
\newblock Electronic properties of single-layer antimony: Tight-binding model,
  spin-orbit coupling, and the strength of effective coulomb interactions.
\newblock {\em Physical Review B}, 95(8):081407, 2017.

\bibitem{Ares2016}
Pablo Ares, Fernando Aguilar-Galindo, David Rodr{\'\i}guez-San-Miguel, Diego~A
  Aldave, Sergio D{\'\i}az-Tendero, Manuel Alcam{\'\i}, Fernando Mart{\'\i}n,
  Julio G{\'o}mez-Herrero, and F{\'e}lix Zamora.
\newblock Mechanical isolation of highly stable antimonene under ambient
  conditions.
\newblock {\em Advanced Materials}, 28(30):6332--6336, 2016.

\bibitem{lloret2019few}
Vicent Lloret, Miguel~{\'A}ngel Rivero-Crespo, Jos{\'e}~Alejandro Vidal-Moya,
  Stefan Wild, Antonio Dom{\'e}nech-Carb{\'o}, Bettina~SJ Heller, Sunghwan
  Shin, Hans-Peter Steinr{\"u}ck, Florian Maier, Frank Hauke, et~al.
\newblock Few layer 2d pnictogens catalyze the alkylation of soft nucleophiles
  with esters.
\newblock {\em Nature communications}, 10(1):1--11, 2019.

\bibitem{Gibaja2016}
C~Gibaja, D~Rodriguez-San-Miguel, P~Ares, J~G{\'o}mez-Herrero, M~Varela,
  R~Gillen, J~Maultzsch, F~Hauke, A~Hirsch, G~Abellan, et~al.
\newblock Angew. chemie-int. ed. 2016, 55, 14345--14349.
\newblock {\em Angew. Chem}, 128:14557--14561, 2017.

\bibitem{zhang2016semiconducting}
Shengli Zhang, Meiqiu Xie, Fengyu Li, Zhong Yan, Yafei Li, Erjun Kan, Wei Liu,
  Zhongfang Chen, and Haibo Zeng.
\newblock Semiconducting group 15 monolayers: a broad range of band gaps and
  high carrier mobilities.
\newblock {\em Angewandte Chemie International Edition}, 55(5):1666--1669,
  2016.

\bibitem{zhang2017}
Shengli Zhang, Wenhan Zhou, Yandong Ma, Jianping Ji, Bo~Cai, Shengyuan~A Yang,
  Zhen Zhu, Zhongfang Chen, and Haibo Zeng.
\newblock Antimonene oxides: emerging tunable direct bandgap semiconductor and
  novel topological insulator.
\newblock {\em Nano letters}, 17(6):3434--3440, 2017.

\bibitem{Lugovskoi2019}
AV~Lugovskoi, MI~Katsnelson, and AN~Rudenko.
\newblock Electron-phonon properties, structural stability, and
  superconductivity of doped antimonene.
\newblock {\em Physical Review B}, 99(6):064513, 2019.

\bibitem{shu2018}
Huabing Shu, Yunhai Li, Xianghong Niu, and JiYuan Guo.
\newblock Electronic structures and optical properties of arsenene and
  antimonene under strain and an electric field.
\newblock {\em Journal of Materials Chemistry C}, 6(1):83--90, 2018.

\bibitem{torbatian2018plasmonic}
Zahra Torbatian and Reza Asgari.
\newblock Plasmonic physics of 2d crystalline materials.
\newblock {\em Applied Sciences}, 8(2):238, 2018.

\bibitem{Prishchenko2018}
DA~Prishchenko, VG~Mazurenko, MI~Katsnelson, and AN~Rudenko.
\newblock Gate-tunable infrared plasmons in electron-doped single-layer
  antimony.
\newblock {\em Physical Review B}, 98(20):201401, 2018.

\bibitem{katsnelson2018}
Guus Slotman, Alexander Rudenko, Edo van Veen, Mikhail~I Katsnelson, Rafael
  Rold{\'a}n, and Shengjun Yuan.
\newblock Plasmon spectrum of single-layer antimonene.
\newblock {\em Physical Review B}, 98(15):155411, 2018.

\bibitem{principi14}
Alessandro Principi, Matteo Carrega, Mark~B. Lundeberg, Achim Woessner, Frank
  H.~L. Koppens, Giovanni Vignale, and Marco Polini.
\newblock Plasmon losses due to electron-phonon scattering: The case of
  graphene encapsulated in hexagonal boron nitride.
\newblock {\em Phys. Rev. B}, 90:165408, 2014.

\bibitem{novko17}
Dino Novko.
\newblock Dopant-induced plasmon decay in graphene.
\newblock {\em Nano Letters}, 17(11):6991, 2017.

\bibitem{ni18}
G.~X. Ni, A.~S. McLeod, Z.~Sun, L.~Wang, L.~Xiong, K.~W. Post, S.~S. Sunku,
  B.-Y. Jiang, J.~Hone, C.~R. Dean, M.~M. Fogler, and D.~N. Basov.
\newblock Fundamental limits to graphene plasmonics.
\newblock {\em Nature}, 557(7706):530--533, may 2018.

\bibitem{baroni2001phonons}
Stefano Baroni, Stefano De~Gironcoli, Andrea Dal~Corso, and Paolo Giannozzi.
\newblock Phonons and related crystal properties from density-functional
  perturbation theory.
\newblock {\em Reviews of Modern Physics}, 73(2):515, 2001.

\bibitem{mostofi2008wannier90}
Arash~A Mostofi, Jonathan~R Yates, Young-Su Lee, Ivo Souza, David Vanderbilt,
  and Nicola Marzari.
\newblock wannier90: A tool for obtaining maximally-localised wannier
  functions.
\newblock {\em Computer physics communications}, 178(9):685--699, 2008.

\bibitem{ponce2016epw}
Samuel Ponc{\'e}, Elena~R Margine, Carla Verdi, and Feliciano Giustino.
\newblock Epw: Electron--phonon coupling, transport and superconducting
  properties using maximally localized wannier functions.
\newblock {\em Computer Physics Communications}, 209:116--133, 2016.

\bibitem{0953-8984-21-39-395502}
Paolo Giannozzi, Stefano Baroni, Nicola Bonini, Matteo Calandra, Roberto Car,
  Carlo Cavazzoni, Davide Ceresoli, Guido~L Chiarotti, Matteo Cococcioni,
  Ismaila Dabo, Andrea~Dal Corso, Stefano de~Gironcoli, Stefano Fabris, Guido
  Fratesi, Ralph Gebauer, Uwe Gerstmann, Christos Gougoussis, Anton Kokalj,
  Michele Lazzeri, Layla Martin-Samos, Nicola Marzari, Francesco Mauri,
  Riccardo Mazzarello, Stefano Paolini, Alfredo Pasquarello, Lorenzo Paulatto,
  Carlo Sbraccia, Sandro Scandolo, Gabriele Sclauzero, Ari~P Seitsonen,
  Alexander Smogunov, Paolo Umari, and Renata~M Wentzcovitch.
\newblock Quantum espresso: a modular and open-source software project for
  quantum simulations of materials.
\newblock {\em Journal of Physics: Condensed Matter}, 21(39):395502, 2009.

\bibitem{Novko2016}
Dino Novko, Marijan {\v{S}}unji{\'c}, and Vito Despoja.
\newblock Optical absorption and conductivity in quasi-two-dimensional crystals
  from first principles: Application to graphene.
\newblock {\em Physical Review B}, 93(12):125413, 2016.

\bibitem{torbatian_PRB}
Zahra Torbatian and Reza Asgari.
\newblock Optical absorption properties of few-layer phosphorene.
\newblock {\em Physical Review B}, 98(20):205407, 2018.

\bibitem{rukelj2016optical}
Zoran Rukelj, Antonio {\v{S}}trkalj, and Vito Despoja.
\newblock Optical absorption and transmission in a molybdenum disulfide
  monolayer.
\newblock {\em Physical Review B}, 94(11):115428, 2016.

\bibitem{despoja2009propagators}
Vito Despoja, Marijan {\v{S}}unji{\'c}, and Leonardo Maru{\v{s}}i{\'c}.
\newblock Propagators and spectra of surface polaritons in metallic slabs:
  Effects of quantum-mechanical nonlocality.
\newblock {\em Physical Review B}, 80(7):075410, 2009.

\bibitem{Caruso2018}
Fabio Caruso, Dino Novko, and Claudia Draxl.
\newblock Phonon-assisted damping of plasmons in three-and two-dimensional
  metals.
\newblock {\em Physical Review B}, 97(20):205118, 2018.

\bibitem{PhysRevB.3.305}
P.~B. Allen.
\newblock Electron-phonon effects in the infrared properties of metals.
\newblock {\em Phys. Rev. B}, 3:305--320, Jan 1971.

\bibitem{Puchkov_1996}
A~V Puchkov, D~N Basov, and T~Timusk.
\newblock The pseudogap state in high- superconductors: an infrared study.
\newblock {\em Journal of Physics: Condensed Matter}, 8:10049, Nov 1996.

\bibitem{xu2017first}
Yuanfeng Xu, Bo~Peng, Hao Zhang, Hezhu Shao, Rongjun Zhang, and Heyuan Zhu.
\newblock First-principle calculations of optical properties of monolayer
  arsenene and antimonene allotropes.
\newblock {\em Annalen der Physik}, 529(4):1600152, 2017.

\bibitem{zhang2017antimonene}
Shengli Zhang, Wenhan Zhou, Yandong Ma, Jianping Ji, Bo~Cai, Shengyuan~A Yang,
  Zhen Zhu, Zhongfang Chen, and Haibo Zeng.
\newblock Antimonene oxides: emerging tunable direct bandgap semiconductor and
  novel topological insulator.
\newblock {\em Nano letters}, 17(6):3434--3440, 2017.

\bibitem{chen2016electronic}
Xianping Chen, Qun Yang, Ruishen Meng, Junke Jiang, Qiuhua Liang, Chunjian Tan,
  and Xiang Sun.
\newblock The electronic and optical properties of novel germanene and
  antimonene heterostructures.
\newblock {\em Journal of Materials Chemistry C}, 4(23):5434--5441, 2016.

\bibitem{zou2016phonon}
Ji-Hang Zou, Zhen-Qiang Ye, and Bing-Yang Cao.
\newblock Phonon thermal properties of graphene from molecular dynamics using
  different potentials.
\newblock {\em The Journal of chemical physics}, 145(13):134705, 2016.

\bibitem{liu2014anisotropic}
Te-Huan Liu, Yin-Chung Chen, Chun-Wei Pao, and Chien-Cheng Chang.
\newblock Anisotropic thermal conductivity of mos2 nanoribbons: Chirality and
  edge effects.
\newblock {\em Applied Physics Letters}, 104(20):201909, 2014.

\bibitem{PhysRevB.91.115412}
Mohammad Elahi, Kaveh Khaliji, Seyed~Mohammad Tabatabaei, Mahdi Pourfath, and
  Reza Asgari.
\newblock Modulation of electronic and mechanical properties of phosphorene
  through strain.
\newblock {\em Phys. Rev. B}, 91:115412, Mar 2015.

\bibitem{esfahani2017superconducting}
Davoud~Nasr Esfahani and Reza Asgari.
\newblock Superconducting critical temperature of hole doped blue phosphorene.
\newblock {\em arXiv preprint arXiv:1710.05554}, 2017.

\bibitem{jain2015strongly}
Ankit Jain and Alan~JH McGaughey.
\newblock Strongly anisotropic in-plane thermal transport in single-layer black
  phosphorene.
\newblock {\em Scientific reports}, 5:8501, 2015.

\bibitem{peng2016low}
Bo~Peng, Hao Zhang, Hezhu Shao, Yuchen Xu, Xiangchao Zhang, and Heyuan Zhu.
\newblock Low lattice thermal conductivity of stanene.
\newblock {\em Scientific reports}, 6:20225, 2016.

\bibitem{kohn1959image}
W~Kohn.
\newblock Image of the fermi surface in the vibration spectrum of a metal.
\newblock {\em Physical Review Letters}, 2(9):393, 1959.

\bibitem{lazzeri2006nonadiabatic}
Michele Lazzeri and Francesco Mauri.
\newblock Nonadiabatic kohn anomaly in a doped graphene monolayer.
\newblock {\em Physical review letters}, 97(26):266407, 2006.

\end{thebibliography}

\end{document}